\documentclass[final,5p,times,twocolumn,authoryear,fleqn]{elsarticle}

\usepackage{amssymb}
\usepackage{amsmath}
\usepackage{hyperref}
\usepackage{mathrsfs}
\usepackage{slashed}
\hypersetup{colorlinks=true}

\journal{Nuclear Physics B}

\def\p{\mbox{\boldmath$\displaystyle\mathbf{p}$}}

\def\e{\mbox{\boldmath$\displaystyle\mathbf{\epsilon}$}}

\def\bv{\mbox{\boldmath$\displaystyle\mathbf{\varphi}$}}

\def\0{\mbox{\boldmath$\displaystyle\mathbf{0}$}}

\def\s{\mbox{\boldmath$\displaystyle\mathbf{\sigma}$}}

\def\x{\mbox{\boldmath$\displaystyle\mathbf{x}$}}
\def\y{\mbox{\boldmath$\displaystyle\mathbf{y}$}}

\newcommand{\dual}[1]{\overset{{}^{{}^{\boldsymbol{\neg}}}}{\smash[t]{#1}}} 
\newcommand{\gdual}[1]{\overset{\:\:{}^{{}^{\boldsymbol{\neg}}}}{\smash[t]{#1}}} 

\begin{document}

\begin{frontmatter}



\title{The Lagrangian for mass dimension one fermionic dark matter}


\author{Cheng-Yang Lee}

\address{Institute of Mathematics, Statistics and Scientific Computation,\\
Unicamp, 13083-859 Campinas, S\~{a}o Paulo, Brazil}
\ead{cylee@ime.unicamp.br}

\begin{abstract}
The mass dimension one fermionic field associated with Elko satisfies the Klein-Gordon but not the Dirac equation. However, its propagator is not a Green's function of the Klein-Gordon operator. We determine the operator in which the associated Green's function is the propagator of the fermionic field. The field is still of mass dimension one, but the obtained Lagrangian resolves the last outstanding issue of mass dimension one fields. This Lagrangian does not admit local gauge invariance. Therefore, the mass dimension one fermions have limited interactions with the Standard Model particles and are natural dark matter candidates.
\end{abstract}


\end{frontmatter}

\section{Introduction}
The theoretical discovery of Elko and the associated mass dimension one fermions by~\cite{Ahluwalia:2004sz,Ahluwalia:2004ab} is a radical departure from the Standard Model (SM). These fermions have renormalizable self-interactions and only interact with the SM particles through gravity and the Higgs boson. These properties make them natural dark matter candidates.

Since their conceptions, Elko and its fermionic fields have been studied in many disciplines. The graviational interactions of Elko have received much attention~\cite{Boehmer:2006qq,Boehmer:2007dh,Boehmer:2007ut,Boehmer:2008rz,Boehmer:2008ah,Boehmer:2009aw,
Boehmer:2010tv,Boehmer:2010ma,
Chee:2010ju,Shankaranarayanan:2009sz,Shankaranarayanan:2010st,Gredat:2008qf,Wei:2010ad,daRocha:2013qhu,
Basak:2012sn,daSilva:2014kfa,Basak:2014qea} while its mathematical properties have been investigated by da Rocha and collaborators~\cite{daRocha:2005ti,
daRocha:2007pz,daRocha:2008we,HoffdaSilva:2009is,daRocha:2009gb,daRocha:2011yr,daRocha:2011xb,
Bernardini:2012sc}. These works established Elko as an inflaton candidate and that it is a flagpole spinor of the Lounesto classification~\cite{Lounesto:2001zz} thus making them fundamentally different from the Dirac spinor. In particle physics, the signatures of these mass dimension one fermions at the Large Hardon Collider have been studied~\cite{Dias:2010aa,Alves:2014kta}. In quantum field theory, much of the attention is focused on the foundations of the construction~\cite{Ahluwalia:2008xi,Ahluwalia:2009rh,Fabbri:2009ka,Fabbri:2009aj,Fabbri:2010va,
Lee:2012thesis,Cavalcanti:2014uta,Nikitin:2014fga}. Their supersymmetric and higher-spin extensions have also been carried out by~\cite{Wunderle:2010yw,Lee:2012td}. An important result is that the fermionic field and its higher-spin generalization violate Lorentz symmetry due to the existence of a preferred direction. This led Ahluwalia and Horvath to suggest that the fermionic field satisfies the symmetry of very special relativity~\cite{Ahluwalia:2010zn,Cohen:2006ky}.

One question remains unanswered in the literature. What is the correct Lagrangian of the mass dimension one fermion? Since the field is constructed using Elko as expansion coefficients which satisfy the Klein-Gordon equation, the naive answer would be the Klein-Gordon Lagrangian. But this has two unsatisfactory aspects. Firstly, the resulting field-momentum anti-commutator is non-local. Secondly, the propagator is not a Green's function of the Klein-Gordon operator.

We determine the Lagrangian for the mass dimension one fermions where these problems are absent. The resulting field is local and the propagator is a Green's function to the operator associated with the field equation. This Lagrangian does not admit local gauge invariance. Therefore, the mass dimension one fermions have limited interactions with the SM particles and are natural dark matter candidates.

\section{The Elko construct}
We briefly review the construction of Elko and its fermionic field. For more details, please refer to the review article~\cite{Ahluwalia:2013uxa}. Elko is a German acronym for \textbf{E}igenspinoren des \textbf{L}adungs\textbf{k}onjugations\textbf{o}perators. They are a complete set of eigenspinors of the charge-conjugation operator of the $(\frac{1}{2},0)\oplus(0,\frac{1}{2})$ representation of the Lorentz group. The charge-conjugation operator is defined as
\begin{equation}
\mathcal{C}=\left(\begin{matrix}
O & -i\Theta^{-1} \\
-i\Theta & O 
\end{matrix}\right)K
\end{equation}
where $K$ complex conjugates anything to its right and $\Theta$ is the spin-half Wigner time-reversal matrix
\begin{equation}
\Theta=\left(\begin{matrix}
0 & -1 \\
1 & 0 
\end{matrix}\right).
\end{equation}
Its action on the Pauli matrices $\s=(\sigma_{1},\sigma_{2},\sigma_{3})$ is
\begin{equation}
\Theta\s\Theta^{-1}=-\s^{*}. \label{eq:Wigner_matrix}
\end{equation}

To construct Elko, we start with $\phi(\e,\sigma)$, a left-handed Weyl spinor in the helicity basis where $\e=\lim_{\mathbf{|p|\rightarrow0}}\hat{\p}$ and
\begin{equation}
\frac{1}{2}\s\cdot\hat{\p}\,\phi(\e,\sigma)=\sigma\phi(\e,\sigma)
\end{equation}
so that $\sigma=\pm\frac{1}{2}$ denotes the helicity. The spinor of arbitrary momentum is obtained by 
\begin{equation}
\phi(\p,\sigma)=\exp\left(-\frac{1}{2}\s\cdot\bv\right)\phi(\e,\sigma)\label{eq:Lboost}
\end{equation}
where  $\bv=\varphi\hat{\p}$ is the rapidity parameter defined as
\begin{equation}
\cosh\varphi=\frac{E_{\mathbf{p}}}{m},\hspace{0.5cm}
\sinh\varphi=\frac{|\p|}{m}.
\end{equation}
Now complex conjugate $\phi(\p,\sigma)$ in eq.~(\ref{eq:Lboost}) and then multiply it from the left by $\Theta$. Apply eq.~(\ref{eq:Wigner_matrix}), we find that $\vartheta\Theta\phi^{*}(\p,\sigma)$ where $\vartheta$ is a phase to be determined, transforms as a right-handed Weyl spinor
\begin{equation}
\vartheta\Theta\phi^{*}(\p,\sigma)=\exp\left(\frac{1}{2}\s\cdot\bv\right)
[\vartheta\Theta\phi^{*}(\e,\sigma)]
\end{equation}
 and has opposite helicity with respect to $\phi(\p,\sigma)$
\begin{equation}
\frac{1}{2}\s\cdot\hat{\p}[\vartheta\Theta\phi^{*}(\p,\sigma)]=
-\sigma[\vartheta\Theta\phi^{*}(\p,\sigma)].
\end{equation}
Putting the right and left-handed Weyl spinor together yields a four-component spinor 
$\chi(\p,\alpha)$ of the form
\begin{equation}
\chi(\p,\alpha)=
\left(\begin{matrix}
\vartheta\Theta\phi^{*}(\p,\sigma)\\
\phi(\p,\sigma)\end{matrix}\right)
\end{equation}
where $\alpha=\mp\sigma$ denotes the dual-helicity nature of the spinor with the top and bottom signs denoting the helicity of the right and left-handed Weyl spinors respectively. The spinor $\chi(\p,\alpha)$ becomes the eigenspinor of the charge-conjugation operator $\mathcal{C}$ with the following choice of phases
\begin{equation}
\mathcal{C}\chi(\p,\alpha)\vert_{\vartheta=\pm i}=\pm\chi(\p,\alpha)\vert_{\vartheta=\pm i}
\end{equation}
thus giving us four Elkos. Spinors with the positive and negative eigenvalues are called the self-conjugate and anti-self-conjugate spinors. They are denoted as
\begin{subequations}
\begin{equation}
\mathcal{C}\xi(\p,\alpha)=\xi(\p,\alpha),\hspace{0.5cm}
\end{equation}
\begin{equation}
\mathcal{C}\zeta(\p,\alpha)=-\zeta(\p,\alpha).
\end{equation}
\end{subequations}
There are subtleties involved in choosing the labellings and phases for the self-conjugate and anti-self-conjugate spinors. The details, including the solutions of the spinors can be found in ~\cite[sec.~II.A]{Ahluwalia:2009rh}. 

The Elko dual which yields the invariant inner-product is defined as~\cite{Speranca:2013hqa,Ahluwalia:2013uxa}
\begin{subequations}
\begin{equation}
\dual{\xi}(\p,\alpha)=[\Xi(\p)\xi(\p,\alpha)]^{\dag}\Gamma,\hspace{0.5cm}\label{eq:Elkodual1}
\end{equation}
\begin{equation}
\dual{\zeta}(\p,\alpha)=[\Xi(\p)\zeta(\p,\alpha)]^{\dag}\Gamma \label{eq:Elkodual2}
\end{equation}
\end{subequations}
where $\dag$ represents Hermitian conjugation and $\Gamma$ is a block-off-diagonal matrix comprised of $2\times2$ identity matrix
\begin{equation}
\Gamma=\left(\begin{matrix}
O & I \\
I & O 
\end{matrix}\right).
\end{equation}
The matrix $\Xi(\p)$ is defined as
\begin{equation}
\Xi(\p)=\frac{1}{2m}\sum_{\alpha}\left[\xi(\p,\alpha)\bar{\xi}(\p,\alpha)-\zeta(\p,\alpha)
\bar{\zeta}(\p,\alpha)\right].
\end{equation}
The bar over the spinors denotes the Dirac dual. The dual ensures that the Elko norms are orthonormal
\begin{equation}
\dual{\xi}(\p,\alpha)\xi(\p,\alpha')=-\dual{\zeta}(\p,\alpha)
\zeta(\p,\alpha')=2m\delta_{\alpha\alpha'}
\end{equation}
and that they satisfy the completeness relation
\begin{equation}
\frac{1}{2m}\sum_{\alpha}\left[\xi(\p,\alpha)\dual{\xi}(\p,\alpha)-\zeta(\p,\alpha)\dual{\zeta}(\p,\alpha)\right]=I.
\end{equation}
The separate spin-sums are
\begin{subequations}
\begin{equation}
\sum_{\alpha}\xi(\p,\alpha)\gdual{\xi}(\p,\alpha)=m[\mathcal{G}(\phi)+I],\label{eq:elko_spinsum1}
\end{equation}
\begin{equation}
\sum_{\alpha}\zeta(\p,\alpha)\gdual{\zeta}(\p,\alpha)=m[\mathcal{G}(\phi)-I]\label{eq:elko_spinsum2}
\end{equation}
\end{subequations}
where $\mathcal{G}(\phi)$ is an off-diagonal matrix
\begin{equation}
\mathcal{G}(\phi)=i\left(\begin{matrix}
0 & 0 & 0 & -e^{-i\phi} \\
0 & 0 & e^{i\phi} & 0 \\
0 & -e^{-i\phi} & 0 & 0\\
e^{i\phi} & 0 & 0 & 0 \end{matrix}\right).
\end{equation}
The angle $\phi$ is defined via the following parametrization of the momentum $\p$
\begin{equation}
\p=|\p|(\sin\theta\cos\phi,\sin\theta\sin\phi,\cos\theta)\label{eq:spc}
\end{equation}
where $0\leq\theta\leq\pi$ and $0\leq\phi<2\pi$.
Multiply eqs.~(\ref{eq:elko_spinsum1}) and (\ref{eq:elko_spinsum2}) with $\xi(\p,\alpha')$ and $\zeta(\p,\alpha')$ from the right and apply the orthonormal relations, we obtain
\begin{subequations}
\begin{equation}
\left[\mathcal{G}(\phi)-I\right]\xi(\p,\alpha)=0,\label{eq:gxi}
\end{equation}
\begin{equation}
\left[\mathcal{G}(\phi)+I\right]\zeta(\p,\alpha)=0. \label{eq:gzeta}
\end{equation}
\end{subequations}
Since these identities have no explicit energy dependence, the corresponding equation in the configuration space have no dynamics and therefore cannot be the field equation for the mass dimension one fermions. Nevertheless, writing the above identities in the configuration space for $\lambda(x)$ is non-trivial and is a task that must be accomplished in order to derive the Hamiltonian. This issue is addressed in the next section.

Identifying the self-conjugate and anti-self-conjugate spinors with the expansion coefficients for particles and anti-particles, the two mass dimension one fermionic fields and their adjoints, with the appropriate normalization are
\begin{subequations}
\begin{align}
\Lambda(x)&=&(2\pi)^{-3/2}\int\frac{d^{3}p}{\sqrt{2mE_{\mathbf{p}}}}\sum_{\alpha}[e^{-ip\cdot x}\xi(\p,\alpha)
a(\p,\alpha)\nonumber\\
&&+e^{ip\cdot x}\zeta(\p,\alpha)b^{\ddag}(\p,\alpha)],
\end{align}
\begin{align}
\dual{\Lambda}(x)&=&(2\pi)^{-3/2}\int\frac{d^{3}p}{\sqrt{2mE_{\mathbf{p}}}}\sum_{\alpha}[e^{ip\cdot x}
\dual{\xi}(\p,\alpha)
a^{\ddag}(\p,\alpha)\nonumber\\
&&+e^{-ip\cdot x}\dual{\zeta}(\p,\alpha)b(\p,\alpha)],
\end{align}
\begin{equation}
\lambda(x)=\Lambda(x)\vert_{b^{\ddag}=a^{\ddag}},
\end{equation}
\begin{equation}
\gdual{\lambda}(x)=\dual{\Lambda}(x)\vert_{b^{\ddag}=a^{\ddag}}.
\end{equation}
\end{subequations}
Here $a(\p,\alpha)$ and $b^{\ddag}(\p,\alpha)$ are the annihilation and creation operators for particles and anti-particles. They satisfy the standard anti-commutation relations
\begin{eqnarray}
\{a(\p',\alpha'),a^{\ddag}(\p,\alpha)\}&=&\{b(\p',\alpha'),b^{\ddag}(\p,\alpha)\}\nonumber\\
&=&\delta_{\alpha'\alpha}\delta^{3}(\p'-\p).
\end{eqnarray}
Note that for the creation operators, we have introduced a new operator $\ddag$ in place of the usual Hermitian conjugation $\dag$. This follows from the observation that since the Dirac and Elko dual are different, it suggests that the corresponding adjoints for the respective particle states may be different. Assuming they are different, it may then become necessary to develop a new formalism for particles states with the new $\ddag$  adjoint in parallel to~\cite{Dirac:1958}. This is an important issue that deserves further study but since it does not affect our objective of deriving the Lagrangian, we shall leave it for future investigation.

\section{The Lagrangian: Defining the problem}\label{3}
There are two reasons why the Klein-Gordon Lagrangian are unsatisfactory for the mass dimension one fermions. Firstly, the field is non-local since the field-momentum anti-commutator is not equal to $i\delta^{3}(\x-\y)I$. Instead, it is given by\footnote{For the rest of the paper, we will be working with $\lambda(x)$, but the results hold for $\Lambda(x)$ also.}
\begin{equation}
\{\lambda(t,\x),\pi_{kg}(t,\y)\}=i\int\frac{d^{3}p}{(2\pi)^{3}}
e^{-i\mathbf{p\cdot(x-y)}}[I+\mathcal{G}(\phi)]\label{eq:lpi}
\end{equation}
where $\pi_{kg}=\partial\gdual{\lambda}/\partial t$ is the conjugate momentum of the Klein-Gordon Lagrangian.
Secondly, the propagator obtained from the fermionic time-order product is
\begin{eqnarray}
S(x,y)&=&\langle\;\vert{T}[\lambda(x)\gdual{\lambda}(y)]\vert\;\rangle \nonumber\\
&=& i\int \frac{d^4 p}{(2\pi)^4} e^{-i p\cdot(x- y)} 
\left[ \frac{I + {\mathcal G}(\phi)}{p\cdot p - m^2 + i\epsilon}\right]. \label{eq:prop}
\end{eqnarray}
This is not a Green's function of the Klein-Gordon operator
\begin{equation}
(\partial^{\mu}\partial_{\mu}+m^{2})S(x,y)=-i\int\frac{d^{4}p}{(2\pi)^{4}}e^{-ip\cdot(x-y)}[I+\mathcal{G}(\phi)]. \label{eq:kgs}
\end{equation}
In eqs.~(\ref{eq:lpi}) and (\ref{eq:kgs}) the problem resides in the three and four-dimensional Fourier transform of $\mathcal{G}(\phi)$ which are non-vanishing thus making the field non-local.\footnote{The Fourier transform of $\mathcal{G}(\phi)$ was first evaluated by J.~Polchinski and was later communicated to me by D.~V.~Ahluwalia.}

\subsection{The inverse of $I+\mathcal{G}(\phi)$}

A consistent field theory should have a well-defined action and transition amplitudes in the operator and the path-integral formalism. Both require the locality of the field and the Lagrangian over the whole space-time continuum. In this respect, it is evident that a Klein-Gordon Lagrangian is inadequate to describe the mass dimension one fermions. Here we derive the Lagrangian that provides a complete description of $\lambda(x)$. We start by determining the operator $\mathcal{O}(x)$ in which the propagator given by eq.~(\ref{eq:prop}) is a Green's function
\begin{equation}
\mathcal{O}(x)S(x,y)=-i\delta^{4}(x-y).
\end{equation}
According to eq.~(\ref{eq:kgs}), this can be achieved by determining the inverse of $I+\mathcal{G}(\phi)$ and the corresponding operator in the configuration space. However this matrix is non-invertible since
\begin{equation}
\det[I+\mathcal{G}(\phi)]=0.
\end{equation}
This problem can be bypassed by considering a more general matrix $I+\tau\mathcal{G}(\phi)$ where $\tau$ is a real constant. Its inverse is
\begin{equation}
[I+\tau\mathcal{G}(\phi)]^{-1}=\frac{I-\tau\mathcal{G}(\phi)}{1-\tau^{2}}.
\label{eq:inverse}
\end{equation}
The singularities can be avoided by taking the limit $\tau\rightarrow\pm1$. This is possible since a simple calculation shows that the inverse is well-defined for all values of $\tau$
\begin{equation}
[I+\tau\mathcal{G}(\phi)]^{-1}[I+\tau\mathcal{G}(\phi)]=
\left(\frac{1-\tau^{2}}{1-\tau^{2}}\right)I=I
\end{equation}
where we have used the identity $\mathcal{G}^{2}(\phi)=I$. 
Therefore, to obtain the inverse of $I+\mathcal{G}(\phi)$, we must first perform a $\tau$-deformation
\begin{equation}
I+\mathcal{G}(\phi)\rightarrow I+\tau\mathcal{G}(\phi).\label{eq:tau}
\end{equation}
The inverse is then given by eq.~(\ref{eq:inverse}).

To determine the form of $[I+\tau\mathcal{G}(\phi)]^{-1}$ in the configuration space, we first need to define an operator $\mathscr{G}$ which corresponds to the matrix $\mathcal{G}(\phi)$ in the configuration space. For this purpose, fractional derivatives must be introduced since the matrix elements of $\mathcal{G}(\phi)$ are proportional to $e^{\pm i\phi}$ and that it can be expressed in terms of complex momenta $p^{\pm}=(p^{1}\pm i p^{2})/\sqrt{2}$ as
\begin{equation}
e^{\pm i\phi}=[p^{\pm}(p^{\mp})^{-1}]^{1/2}.
\end{equation}
There are many definitions of fractional derivatives. The fractional derivative that is appropriate for our task is the Fourier fractional derivative~\cite[pg~562]{Fourier}~\cite{Herrmann:2011zza}. The general properties of fractional derivatives, including Fourier's definition, are given in app.~\ref{A}. The operator $\mathscr{G}$ is defined as
\begin{equation}
\mathscr{G}=i\left(\begin{matrix}
0 & 0 & 0 & -\partial^{1/2}_{+}\partial^{-1/2}_{-} \\
0 & 0 & \partial^{-1/2}_{+}\partial^{1/2}_{-} & 0 \\
0 & -\partial^{1/2}_{+}\partial^{-1/2}_{-} & 0 & 0  \\
\partial^{-1/2}_{+}\partial^{1/2}_{-} & 0 & 0 & 0
\end{matrix}\right)
\end{equation}
where $\partial_{\pm}$ act on the complex coordinates $x^{\pm}=(x^{1}\pm ix^{2})/\sqrt{2}$.


An interesting property of the Fourier fractional derivative, which turns out to be important for our construct is its non-uniqueness. To see what this means, let us consider $i\partial_{+}^{1/2}\partial_{-}^{-1/2}f (x)$, an element of $\mathscr{G}f(x)$ where $f(x)$ is a test function with the Fourier transform
\begin{equation}
f(x)=\int \frac{d^{3}p}{(2\pi)^{3}} e^{-ip\cdot x}F(\p).
\end{equation}
Using eq.~(\ref{eq:fdf}), we obtain
\begin{equation}
\partial_{-}^{-1/2}f (x)
=\int \frac{d^{3}p}{(2\pi)^{3}} e^{-ip\cdot x}e^{-i\pi/4}e^{-in_{-}\pi}p_{+}^{-1/2}F(\p)
\end{equation}
where we have used the fact that $i^{-1/2}=e^{-i\pi/4}e^{-in_{-}\pi}$ has two roots with $n_{-}=0,1$. Acting on the expression again with $i\partial_{+}^{1/2}$ we obtain
\begin{equation}
i\partial_{+}^{1/2}\partial_{-}^{-1/2}f(x)
=\int \frac{d^{3}p}{(2\pi)^{3}}\; (i\omega)
e^{-ip\cdot x}e^{-i\phi}F(\p)
\end{equation}
where $\omega=e^{i(n_{+}-n_{-})\pi}$
with $n_{\pm}=0,1$.
Due to the ambiguity of the phases, in order to obtain $\mathcal{G}(\phi)$ in the momentum space, all the elements of $\mathscr{G}f(x)$ must have the same phases so that
\begin{equation}
\mathscr{G}f (x)
=\int \frac{d^{3}p}{(2\pi)^{3}}\omega e^{-ip\cdot x}\mathcal{G}(\phi)F(\p).\label{eq:gf}
\end{equation}

For functions such as $f(x)$ comprised of a single Fourier transform, $\omega$ is a global phase and its value is unimportant. But since $\lambda(x)$ is a sum of the Fourier transform of the self-conjugate and anti-self-conjugate spinors, the action of $\mathscr{G}$ on the field yields 
\footnote{In this paper, we assume that the fermionic field $\lambda(x)$ and its dual $\gdual{\lambda}(x)$, both of which are a sum of the Fourier transform of the self-conjugate and anti-self-conjugate spinors and its dual, are well-defined.}
\begin{align}
\mathscr{G}\lambda(x)&=&(2\pi)^{-3}\int\frac{d^{3}p}{\sqrt{2mE_{\mathbf{p}}}}\sum_{\alpha}
[e^{-ip\cdot x}\omega_{\xi}\xi(\p,\alpha)a(\p,\alpha)\nonumber\\
&&-e^{ip\cdot x}\omega_{\zeta}\zeta(\p,\alpha)a^{\ddag}(\p,\alpha)].\label{eq:gl}
\end{align}
Since the phases can take the values of $\pm1$, $\mathscr{G}\lambda(x)$ has four possible solutions. Out of the four possibilities, as we will show in the next section, the only solution which yields a positive-definite free Hamiltonian is
\begin{equation}
\omega_{\xi}=-\omega_{\zeta}=1. \label{eq:ps}
\end{equation}
This then gives us the equation
\begin{equation}
\mathscr{G}\lambda(x)=\lambda(x).
\end{equation} 
which is the counterpart of eqs.~(\ref{eq:gxi}) and (\ref{eq:gzeta}) in the configuration space.

Having defined $\mathscr{G}$, the inverse of $I+\tau\mathcal{G}(\phi)$ in the configuration space is given by
\begin{equation}
\mathcal{A}=\frac{I-\tau\,\mathscr{G}}{1-\tau^{2}}
\end{equation}
Now we apply $\mathcal{A}$ to eq.~(\ref{eq:kgs}). The $\tau$-deformed propagator is
\begin{equation}
S^{(\tau)}(x,y)=i\int \frac{d^4 p}{(2\pi)^4} e^{-i p\cdot(x- y)} 
\left[ \frac{I + \tau{\mathcal G}(\phi)}{p\cdot p - m^2 + i\epsilon}\right].\label{eq:tau_prop}
\end{equation}
When $\mathcal{A}$ acts on the $S^{(\tau)}(x,y)$, we must take $\omega=1$ so that in the limit $\tau\rightarrow1$, we obtain
\begin{equation}
\lim_{\tau\rightarrow1}\mathcal{A}(\partial^{\mu}\partial_{\mu}+m^{2})S^{(\tau)}(x,y)=-i\delta^{4}(x-y)I. \label{eq:green}
\end{equation}
The operator $\mathcal{O}(x)$, in which the propagator is a Green's function of is therefore
\begin{equation}
\mathcal{O}(x)=\mathcal{A}(\partial^{\mu}\partial_{\mu}+m^{2}).\label{eq:operator}
\end{equation}

Based on the form of $\mathcal{O}(x)$, the Lagrangian for $\lambda(x)$ is
\begin{equation}
\mathscr{L}=\mathcal{A}_{ab}
(\partial^{\mu}\gdual{\lambda}_{a}\partial_{\mu}\lambda_{b}-m^{2}\gdual{\lambda}_{a}\lambda_{b})
\label{eq:Lagrangian}
\end{equation}
where we sum over the repeated indices. The operator $\mathcal{A}$ is dimensionless so $\lambda(x)$ remains a mass dimension one field. The field equation is
\begin{equation}
\mathcal{A}_{ab}(\partial^{\mu}\partial_{\mu}+m^{2})\lambda_{b}=0.
\end{equation}
The operator $\mathcal{A}$ does not affect the solutions of the field equation but it ensures the locality of the field and hence the equivalence between the operator and the path-integral formalism. We now show that the field is local and that the `inverse' of the action yields the propagator given by eq.~(\ref{eq:prop}).


\subsection{Locality structure} \label{locality}

The locality of $\lambda(x)$ is determined by $\{\lambda(t,\x),\lambda(t,\y)\}$,
$\{\pi(t,\x),\pi(t,\y)\}$ and $\{\lambda(t,\x),\pi(t,\y)\}$. The first anti-commutator identically vanishes~\cite[sec.~III.B]{Ahluwalia:2009rh}. When the Lagrangian is Klein-Gordon, the second anti-commutator identically vanishes and the third is given by eq.~(\ref{eq:lpi}). We show that the field associated with eq.~(\ref{eq:Lagrangian}) is local by computing the relevant anti-commutators. The conjugate momentum is
\begin{equation}
\pi_{b}(y)=\mathcal{A}_{ab}(y)\frac{\partial\gdual{\lambda}_{a}}{\partial t}(y)
\end{equation}
which differs from $\pi_{kg}(y)$ by a factor of $\mathcal{A}(y)$. Since fractional derivatives commute, the anti-commutator between the conjugate momentum is
\begin{equation}
\{\pi(t,\x),\pi(t,\y)\}=\mathrm{O}.
\end{equation}
A direct evaluation of the $\tau$-deformed field-momentum anti-commutator now yields
\begin{equation}
\lim_{\tau\rightarrow1}\{\lambda(t,\x),\pi(t,\y)\}^{(\tau)}=i\delta^{3}(\x-\y)I.
\end{equation}

To prove the equivalence between the operator and the path-integral formalism, we derive the propagator using path-integral by inverting the action and show that it is identical to eq.~(\ref{eq:prop}). The action for $\lambda(x)$ is
\begin{eqnarray}
\mathcal{I}&=&
\int d^{4}x\;\mathcal{A}_{ab}(\partial^{\mu}\gdual{\lambda}_{a}\partial_{\mu}\lambda_{b}-m^{2}\gdual{\lambda}_{a}\lambda_{b})\nonumber\\
&=&\int d^{4}x d^{4}y\; \mathscr{D}_{ab}(y,x)\gdual{\lambda}_{a}(y)\lambda_{b}(x).
\end{eqnarray}
To determine $\mathscr{D}_{ab}(x,y)$, we rewrite the action as
\begin{align}
\mathcal{I}
=\int d^{4}x d^{4}y\; 
\left[\mathcal{A}_{ab}(x)\left(\frac{\partial}{\partial y^{\mu}}\frac{\partial}{\partial x_{\mu}}-m^{2}\right)\delta^{4}(x-y)\right]\gdual{\lambda}_{a}(y)\lambda_{b}(x)
\end{align}
where we have used the identity
\begin{align}
\int d^{4}x \mathcal{A}_{ab}(x)[\delta^{4}(x-y)\lambda_{b}(x)]=
\int d^{4}x[\mathcal{A}_{ab}\delta^{4}(x-y)]\lambda_{b}(x).\label{eq:identity}
\end{align}
The operator $\mathscr{D}(x,y)$ is then given by
\begin{equation}
\mathscr{D}(x,y)=
\int\frac{d^{4}p}{(2\pi)^{4}}e^{-ip\cdot (x-y)}
\left[\frac{I-\omega\tau\mathcal{G}(\phi)}{1-\tau^{2}}\right](p\cdot p-m^{2})
\end{equation}
where $\omega=\pm1$ arises from the action of $\mathcal{A}$ on the Dirac-delta function. The phase is determined by noting that out of the two possibilities, only $\omega=1$ yields the correct result, namely
\begin{equation}
i\mathscr{D}^{-1}(x,y)=i\int\frac{d^{4}p}{(2\pi)^{4}}e^{-ip\cdot(x-y)}
\frac{I+\tau\mathcal{G}(\phi)}{p\cdot p-m^{2}+i\epsilon},
\end{equation}
so that in the limit $\tau\rightarrow1$, it is in agreement with eq.~(\ref{eq:prop}).


For the final part of this section, we show that the free Hamiltonian is positive-definite. This is achieved by showing that the free Hamiltonian, as a function of $\lambda(x)$ and $\gdual{\lambda}(x)$ is identical to the one given in~\cite[eq.~(97)]{Lee:2012td}. In order to obtain the correct Hamiltonian, we must take the conjugate momentum associated with both $\lambda(x)$ and $\gdual{\lambda}(x)$ into account where the later is defined as
\begin{equation}
\dual{\pi}_{a}=-\frac{\partial\mathscr{L}}{\partial\gdual{\lambda}_{a}/\partial t}=
-\mathcal{A}_{ab}\frac{\partial\lambda_{b}}{\partial t}.
\end{equation}
The Hamiltonian is then given by the Legendre transformation
\begin{equation}
H=\int d^{3}x\left[\frac{\partial\gdual{\lambda}_{a}}{\partial t}
\left(\mathcal{A}_{ab}\frac{\partial\lambda_{b}}{\partial t}\right)
-\left(\mathcal{A}_{ab}\frac{\partial\lambda_{b}}{\partial t}\right)
\frac{\partial\gdual{\lambda}_{a}}{\partial t}-\mathscr{L}\right].
\end{equation}
In obtaining the first term, we have used the definition of $\mathcal{A}$ and the integration by parts rule for the Fourier fractional derivative. The first two terms can be simplified further since in the limit $\tau\rightarrow1$, we have a simple identity
\begin{equation}
\lim_{\tau\rightarrow1}\mathcal{A}_{ab}\lambda_{b}(x)=\frac{1}{2}\lambda_{a}(x).
\end{equation}
Using the identity
\begin{eqnarray}
&&\hspace{-0.8cm}\int d^{3}x\, \mathcal{A}_{ab}(\partial^{\mu}\gdual{\lambda}_{a}\partial_{\mu}\lambda_{b}-m^{2}\gdual{\lambda}_{a}\lambda_{b})\nonumber\\
&&\hspace{0.5cm}=\frac{1}{2}
\int d^{3}x\, (\partial^{\mu}\gdual{\lambda}_{a}\partial_{\mu}\lambda_{a}-m^{2}\gdual{\lambda}_{a}\lambda_{a}),
\end{eqnarray}
the Hamiltonian simplifies to
\begin{align}
H=\frac{1}{2}\int d^{3}x \left(-\frac{\partial\lambda_{a}}{\partial t}
\frac{\partial\gdual{\lambda}_{a}}{\partial t}-\partial_{i}\gdual{\lambda}_{a}\partial^{i}\lambda_{a}+m^{2}\gdual{\lambda}_{a}\lambda_{a}\right).
\end{align}
This is identical to~\cite[eq.~(97)]{Lee:2012td} and is therefore positive-definite.


\section{A natural dark matter candidate}

The Lagrangian derived in this paper provides a simple explanation to the limited SM interactions. The Lagrangian, with the presence of fractional derivative, does not admit local gauge invariance. To see this, consider a local $U(1)$ transformation. Using the Leibniz rule for fractional derivative given by eq.~(\ref{eq:fLeibniz}), we obtain
\begin{equation}
\partial_{\pm}^{\alpha}[e^{iq\gamma(x)}\lambda(x)]=\sum_{j=0}^{\infty}
\left(\begin{matrix}
\alpha \\
j
\end{matrix}\right)
[\partial_{\pm}^{\alpha-j}e^{iq\gamma(x)}][\partial_{\pm}^{j}\lambda(x)].\label{eq:u1}
\end{equation}
Equation~(\ref{eq:u1}) makes it evident that there are no covariant fractional derivatives under both local Abelian and non-Abelian gauge transformations. The best one could do is linear gauge invariance, demanding the Lagrangian to be invariant under $e^{iq\gamma(x)}\sim 1+iq\gamma(x)+O(q^{2})$ for small $q$~\cite{Herrmann:2007xq,Herrmann:2011zza}. In our opinion, this is unsatisfactory. Local gauge-invariance, if not broken, should be an exact symmetry. It follows that the remaining possibilities are self-interactions, interactions with scalar fields and gravity. Note however, that the simplest interactions of the form $g(\gdual{\lambda}\lambda)^{2}$, $h\gdual{\lambda}\lambda\phi$ and $h'\gdual{\lambda}\lambda\phi^{2}$ are incompatible with the free Lagrangian. The incompatibility arises from the fact that the counter-terms generated by the free Lagrangian which contains fractional derivative, will not be able to absorb all the divergent terms generated by the mentioned interactions. The resolution to this problem is to introduce appropriate interactions containing fractional derivatives. This is beyond the scope of this paper and will be investigated elsewhere.

In the context of the SM, the lack of local gauge invariance means that the mass dimension one fermions can only interact with the SM particles through gravity and the Higgs boson thus making them natural dark matter candidates. 

\section{Conclusions}
%

The Lagrangian derived in this paper addressed the remaining outstanding problems of the free mass dimension one fermions. It not only preserves the mass dimensionality and renormalizable self-interaction, the field is now local and the propagator is a Green's function to the operator given in eq.~(\ref{eq:operator}). At the same time, the lack of local gauge invariance of the Lagrangian strengthens the claim that the mass dimension one fermions are dark matter candidates.

While the mass dimension one fermionic field violates Lorentz symmetry, in light of the new Lagrangian, it is nevertheless a well-defined quantum field in the sense that it has a positive-definite free Hamiltonian, it is local and furnishes fermionic statistics. These properties are highly non-trivial. They require careful choices of expansion coefficients, adjoints and a Lagrangian with fractional derivatives. These results strongly suggest that the mass dimension one fermions have a well-defined space-time symmetry. Additionally, they also suggest the existence of a new class of interacting quantum field theory where fractional calculus plays an important role.

 If the mass dimension one fermionic fields are invariant under very special relativity as proposed by Ahluwalia and Horvath, the effects of Lorentz violation would be minimal since very special relativity is compatible with the null results of the Michelson-Morley experiments and other well-known relativistic effects~\cite{Ahluwalia:2010zn,Horvath}. We would expect discrete symmetry violations and scattering cross-sections involving mass dimension one fermions to have dependence on a preferred direction. 
 
On a broader picture, this theoretical construct presents an interesting new paradigm. Space-time symmetry may be a mere reflection of the symmetry of the rods and clocks comprised of the SM particles. Space-time, according to the rods and clocks made of dark matter, may paint a completely different picture.

\appendix

\section{Fractional derivatives}\label{A}
All fractional derivatives satisfy the following properties. Let $\alpha$ be an arbitrary real number, in the limit $\alpha\rightarrow n$ where $n$ is a positive integer,
\begin{equation}
\lim_{\alpha\rightarrow n}\frac{d^{\alpha}}{dx^{\alpha}}f(x)=\frac{d^{n}}{dx^{n}}f(x).
\end{equation}
The operations are linear
\begin{subequations}
\begin{equation}
\frac{d^{\alpha}}{dx^{\alpha}}[cf(x)]=c\frac{d^{\alpha}}{dx^{\alpha}}f(x),
\end{equation}
\begin{equation}
\frac{d^{\alpha}}{dx^{\alpha}}[f(x)+g(x)]=
\frac{d^{\alpha}}{dx^{\alpha}}f(x)+\frac{d^{\alpha}}{dx^{\alpha}}g(x).
\end{equation}
\end{subequations}
The Leibniz rule is
\begin{equation}
\frac{d^{\alpha}}{dx^{\alpha}}(fg)=\sum_{j=0}^{\infty}
\left(\begin{matrix}
\alpha \\
j
\end{matrix}\right)
\left(\frac{d^{\alpha-j}}{dx^{\alpha}}f\right)\left(\frac{d^{j}}{dx^{j}}g\right)\label{eq:fLeibniz}
\end{equation}
where the binomial coefficient is generalized to arbitrary real numbers by the $\Gamma(\alpha)$ function
\begin{equation}
\left(\begin{matrix}
\alpha \\
j
\end{matrix}\right)
=\frac{\Gamma(\alpha+1)}{\Gamma(j+1)\Gamma(\alpha-j+1)}.
\end{equation}

The Fourier fractional derivative is defined as follows. Let $f(x)$ and $F(k)$ be two single-variable functions related by the Fourier transform
\begin{subequations}
\begin{equation}
f(x)=\frac{1}{2\pi}\int_{-\infty}^{\infty}dk\, e^{-ikx}F(k),
\end{equation}
\begin{equation}
F(k)=\frac{1}{2\pi}\int_{-\infty}^{\infty}dx\,e^{ikx}f(x).
\end{equation}
\end{subequations}
The definition of the fractional derivative on $f(x)$ is a straightforward generalisation of the usual derivative 
\begin{equation}
\frac{d^{\alpha}}{dx^{\alpha}}f(x)=\frac{1}{2\pi}\int_{-\infty}^{\infty}dk\, (-ik)^{\alpha}
e^{-ikx}F(k).\label{eq:fdf}
\end{equation}
This formula is valid for all values of $\alpha$. The only conditions needed are the existence of the Fourier transform for $f(x)$ and its fractional derivative. Using eq.~(\ref{eq:fdf}), we obtain the integration by parts rule
\begin{equation}
\int dx \,\left[f(x)\frac{d^{\alpha}}{dx^{\alpha}}g(x)\right]=
(-1)^{\alpha}\int dx \,\left[g(x)\frac{d^{\alpha}}{dx^{\alpha}}f(x)\right].\label{eq:ibp}
\end{equation}

\section*{Acknowledgements}

I would like to thank N.~Faustino and G.~S.~de Souza for discussions at the early stage of this work. I am grateful to R.~da~Rocha for suggestions and reading the initial manuscript. During the revision of the manuscript, I have benefited greatly from numerous discussions with D.~V.~Ahluwalia. Without his insistence on the importance of phases and constant encouragement, this work would not have been possible. This research is supported by the CNPq grant 313285/2013-6.







\end{document}